\documentclass[twocolumn,showpacs,amsmath,amssymb,superscriptaddress,prl]{revtex4}
\usepackage{graphicx}
\usepackage{dcolumn}
\usepackage{mathrsfs}
\usepackage{bm}
\begin{document}

\preprint{}


\title{ Experimental Observation of Simultaneous Wave and Particle Behaviors in a Narrowband Single Photon's Wave Packet}

\author{Hui Yan}
\email{yanhui@scnu.edu.cn}
\affiliation{Laboratory of Quantum
Engineering and Quantum Materials, School of Physics and
Telecommunication Engineering, South China Normal University,
Guangzhou 510006, China}

\author{Kaiyu Liao}
\affiliation{Laboratory of Quantum Engineering and Quantum
Materials, School of Physics and Telecommunication Engineering,
South China Normal University, Guangzhou 510006, China}

\author{Zhitao Deng}
\affiliation{Laboratory of Quantum Engineering and Quantum
Materials, School of Physics and Telecommunication Engineering,
South China Normal University, Guangzhou 510006, China}

\author{Junyu He}
\affiliation{Laboratory of Quantum Engineering and Quantum
Materials, School of Physics and Telecommunication Engineering,
South China Normal University, Guangzhou 510006, China}

\author{Zheng-Yuan Xue}
\affiliation{Laboratory of Quantum Engineering and Quantum
Materials, School of Physics and Telecommunication Engineering,
South China Normal University, Guangzhou 510006, China}

\author{Zhi-Ming Zhang}
\affiliation{Laboratory of Quantum Engineering and Quantum
Materials, School of Information and Photoelectronic Science and
Engineering, South China Normal University, Guangzhou 510006,
China}

\author{Shi-Liang Zhu}
\email{slzhu@nju.edu.cn}
 \affiliation{National Laboratory of
Solid State Microstructures, School of Physics, Nanjing
University, Nanjing 210093, China}
\affiliation{Laboratory of
Quantum Engineering and Quantum Materials, School of Physics and
Telecommunication Engineering, South China Normal University,
Guangzhou 510006, China}


\begin{abstract}
 Light's wave-particle duality is at the heart of quantum mechanics
and can be well illustrated by Wheeler's delayed-choice
experiment. The choice of inserting or removing the second
classical (quantum) beam splitter in a Mach-Zehnder interferometer
determines the classical (quantum) wave-particle behaviors of a
photon. In this paper, we report our experiment using the
classical beam splitter to observe the simultaneous wave-particle
behaviors in the wave-packet of a narrowband single photon. This
observation suggests that it is necessary to generalize the
current quantum wave-particle duality theory.  Our experiment
demonstrates that the produced wave-particle state can be
considered  an additional degree of freedom and can be utilized in
encoding quantum information.
\end{abstract}

\pacs{03.65.Ta, 42.50.Dv,  03.67.-a, 42.50.Ex}

\maketitle

The dual wave-particle nature of light has been debated for
centuries\cite{1,2}. In order to test the wave-particle duality,
Wheeler proposed the famous delayed-choice gedanken
experiment\cite{3, 4}. The main experimental setup is a Mach-Zehnder
interferometer. After a photon passes through the first beam
splitter (BS), the choice of inserting or removing the second BS is
then randomly determined. If the second BS is present, the photon
travels through both beams of the Mach-Zehnder interferometer and
interference fringes can be observed, indicating the wave behavior
of the photon. If the second BS is absent, the photon randomly
travels through one beam of the Mach-Zehnder interferometer and only
one of the two output ports has a click, showing the full
"which-path" information and the particle properties of the photon.
Since Wheeler's proposal, many delayed-choice experiments have been
realized\cite{5,6,7,8,gao,9}. All of the results support Bohr's
original complementary principle\cite{2} (classical wave-particle
duality), which states that a photon may behave either as a particle
or a wave, depending on the measurement setup, but the two aspects,
particle and wave,  appear to be incompatible and are never observed
simultaneously.

Very recently, Ionicioiu and Terno proposed a quantum version of
the above delayed-choice experiment, in which a quantum BS (which
can be simultaneously present and absent) is utilized to replace
the second BS in the Mach-Zehnder interferometer\cite{10,s}. In
this case a continuous morphing behavior between wave and particle
is expected. Soon after, several experiments were conducted, and
their results confirm the morphing behavior between wave and
particle\cite{11,12,13}. These experiments suggest that a naive
"wave or particle" description (i.e., the classical wave-particle
duality) of light  is inadequate and the generalization to a
quantum version is necessary. In this quantum version, light
exhibits particle- or wave-like behavior depending on the
experimental apparatus: with a quantum detecting setup, light can
simultaneously behave as a particle and as a wave, whereas with a
classical setup, light behaves as a particle or as a
wave\cite{11}.

In this work, by using heralded narrowband single photons, we
experimentally observe the quantum wave-particle behaviors using a
classical setup, the same detecting setup used in Ref.\cite{9} to
realize Wheeler's classical delayed-choice experiment. Benefiting
from the long temporal length, we simultaneously and directly
observe the wave and particle behaviors in a single photon's wave
packet by classically inserting or removing the second BS when
part of the wave packet passes through the second BS. Our results
suggest that a further generalization of the aforementioned
quantum wave-particle duality is necessary.
One can create the superposition state of the wave and particle,
and then observe its quantum behaviors with a classical detecting
setup. In this aspect, the wave-particle character of the photon
can be considered  an additional degree of freedom, in much the
same way as polarization, spin, momentum, and so on \cite{14}.
Moreover, the approach developed in our experiment is a convenient
way to manipulate the wave-particle state, which has exciting
potential given that the wave-particle superposition state can be
used as a qubit for quantum information processing\cite{14}.
Therefore our work  will not only be useful for understanding  the
wave-particle duality and Bohr's complementarity principle but
also open up the possibility of directly using wave-particle state
in encoding information.

 Our
experimental setup is shown in Fig. 1. A heralded narrowband single
photon with the coherence time $\tau$ around 400 ns
\cite{15,16,17,18,19,shibaosen} is sent through a polarization BS
enabled Mach-Zehnder interferometer. The detail of the heralded
narrowband single photon source is described in Supplemental
Material\cite{Supplemental}, which has also been described in our
previous papers \cite{19,cpl}. The photon guided by a single mode
fibre from the source is equally split by BS$_{\text{in}}$ into two
spatially separated paths $|0\rangle$ and $|1\rangle$ associated
with orthogonal polarizations. Then the initial photon state becomes
the superposition $(|0\rangle+|1\rangle)/\sqrt{2}$. The phase shift
$\varphi$ between the two interferometer arms is varied by a
piezoelectric transducer, resulting in the state
$|\Psi\rangle=(|0\rangle+ e^{i\varphi} |1\rangle)/\sqrt{2}$. Both
modes are then recombined on a second controllable 'BS'
(BS$_{\text{out}}$, see Supplemental Material\cite{Supplemental}),
which consists of two BS (BS$^1_{\text{out}}$ and
BS$^2_{\text{out}}$) and an electro-optical modulator (EOM), before
a final measurement in the logical $\{ |0\rangle,|1\rangle\}$ basis.
The inserting or removing of this controllable 'BS$_{\text{out}}$'
is determined when part of the narrowband photon's wave packet
passes through.

\begin{widetext}

\begin{figure}
\begin{center}
\includegraphics[width=12cm]{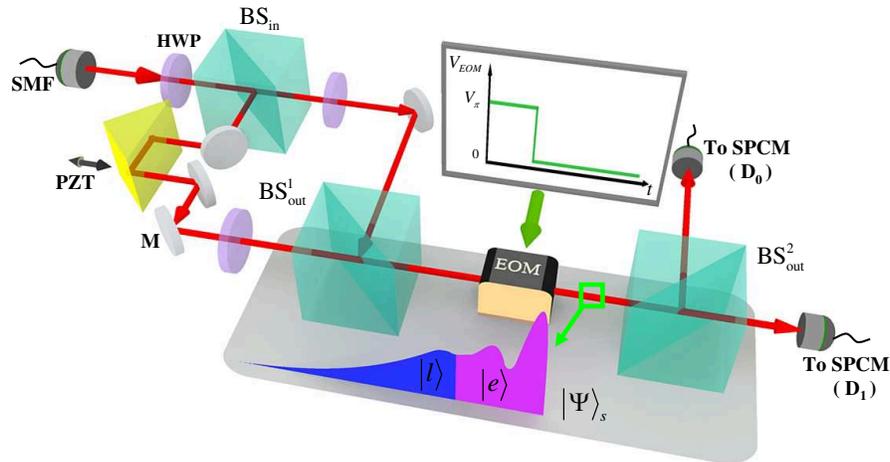}\label{fig:Configuration}
\caption{Experiment setup. The main setup is a polarization beam
splitter (BS) enabled Mach-Zehnder interferometer. A photon is
split by BS$_{\text{in}}$ into two modes with different orthogonal
polarizations and then spatially recombined by
BS$_{\text{out}}^1$. An electro-optical modulator (EOM) is used as
a controllable half-wave plate. BS$_{\text{out}}^2$,  positioned
after the EOM, is used to mix the two orthogonal polarizations or
split the two modes. When $V_{\pi}$ voltage is applied to the EOM,
the interference configuration is closed and the interference
fringe will be detected at the two output ports. When no voltage
is applied, the interferometer configuration is open, and each of
the two output ports provides the full "which-path" information.
Note, M: Mirror; SMF: single mode fibre; HWP: half wave plate;
SPCM: single photon count module; PZT: piezo-electric transducer;
$D_{0,1}$: detectors; $|e\rangle$, $|l\rangle$: time-bin degree
states; $|\Psi\rangle_{s}$: single photon wave function.}
\end{center}
\end{figure}

\end{widetext}

Because of the long temporal length of the single narrowband
photon, the time-bin degree of freedom can be well manipulated and
the time-bin information can be easily
detected\cite{Brendel,Donohue,YanCPL}. We can split the photon
into the  time-bin superposition
$|\Psi\rangle=\cos\alpha|e\rangle+e^{i\gamma}\sin\alpha|l\rangle$,
where $|e\rangle$ $(|l\rangle)$ is the early (late) time bin. The
two modes are first overlapped by BS$^1_\text{out}$ but can still
be identified by their polarization. Then the choice between
closed or open interferometer configuration is achieved with the
EOM: either no voltage or $V_\pi$ voltage is applied to the EOM.
In our experiment, $V_\pi$ is applied to the EOM at the early time
bin.
Therefore,
the BS$_\text{out}$ is present and the interferometer is  closed
at the $|e\rangle$ time slot. In this case, the statistics of
measurements at both detectors $D_{0}$ and $D_{1}$ will depend on
the phase $\varphi$, which will reveal the wave nature of the
photon. Thus the photon is in the 'wave' state given by
\begin{eqnarray}
|\Psi\rangle_{\text{w}}=|e\rangle\otimes|\psi\rangle_{\text{w}},\
\ |\psi\rangle_{\text{w}}=
\cos\frac{\varphi}{2}|0\rangle-\emph{i}\sin\frac{\varphi}{2}|1\rangle.
\label{eq:pbs21}
\end{eqnarray}
On the other hand, no voltage is applied  at the late time bin and
thus the controllable BS$_\text{out}$ isn't present for the
$|l\rangle$ time slot; hence, the interferometer is left open. In
this case, both detectors will click with equal probability, which
will reveal the particle nature of the photon. The photon is in
the 'particle' state given by
\begin{eqnarray}
|\Psi\rangle_{\text{p}}=|l\rangle\otimes |\psi\rangle_{\text{p}},
\ \
|\psi\rangle_{\text{p}}=\frac{1}{\sqrt{2}}(|0\rangle+e^{i\varphi}|1\rangle).
\label{eq:pbs22}
\end{eqnarray}
Therefore, the main feature of the EOM is that it can split the
single photon into two  time bins and  further entangle the time
bin degree of freedom with the wave-particle state. For the
superposition state in the time bin degree, the global state
$|\Psi\rangle_{s}$ of the system after the EOM becomes
\begin{eqnarray}
|\Psi\rangle_{s}(\alpha,\varphi,\gamma)=\cos\alpha|\Psi\rangle_{\text{w}}
+e^{i\gamma}\sin\alpha|\Psi\rangle_{\text{p}}.
\label{Psi}
\end{eqnarray}
After BS$^2_\text{out}$, we measure the photon state, and the
probability of detecting the photon at detector $D_{0}$ is then
given by
\begin{eqnarray}
\label{ID0}
I_{D_{0}}(\alpha,\varphi)=\cos^{2}\left(\frac{\varphi}{2}\right)\cos^{2}\alpha+\frac{1}{2}\sin^{2}\alpha,
\end{eqnarray}
whereas intensity at $D_{1}$ is  $I_{D_{1}}=1-I_{D_{0}}$.

Although the experimental setup shown in Fig.1 is the same as that
in the classical delayed-choice experiment\cite{9}, the statistics
are quite different, which should be explained with the
wave-particle superposition state as in the quantum delayed-choice
experiment\cite{10}. Compared with the experiment in Ref.
\cite{9}, the main difference in our experiment is that a heralded
single photon with long temporal length is used, thus the single
photon can  have three degree of freedoms: time-bin,
wave-particle, and polarization. Furthermore, the time bin states
$|e\rangle$ and $|l\rangle$ are entangled with the wave-particle
state $|\psi\rangle_\text{w}$ and $|\psi\rangle_\text{p}$. Hence,
by choosing the proper measurement basis (different time slots for
$|e\rangle$ and $|l\rangle$) which can be controlled by EOM, we
can demonstrate the quantum character of the wave-particle duality
in this classical setup according to the above theoretical
analysis.

\begin{widetext}

\begin{figure}
\begin{center}
\includegraphics[width=12cm]{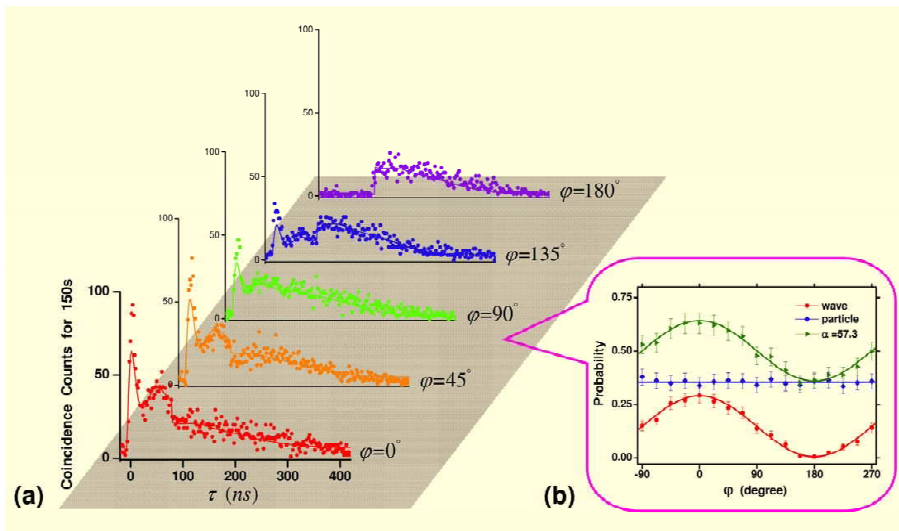}\label{fig:2}
\caption{Temporal wave packet of the single photon measured
through coincidence counts. (a) The total temporal length of the
single photon is about 400 ns. During the first 80 ns, the voltage
$V_{\pi}$ is applied to the EOM, and a clear interference fringe
is observed along with the change of the phase difference
$\varphi$ from $0^{\circ}$ to $180^{\circ}$. After 80 ns, no
voltage is applied to the EOM, and the coincidence counts remain
stable with different $\varphi$. (b) Probability of detecting a
photon with $D_{0}$ during different time slots. The red line with
filled circles stands for $0 \sim 80$ ns. The blue
 line with prismatic plots  stands for $80 \sim 400$ ns. The green
line with triangular plots stands for $0 \sim 400$ ns. For both
(a) and (b), the plots and lines are experimental data and
theoretically predictions, respectively.}
\end{center}
\end{figure}

\end{widetext}

 As a
typical example of experimental results, the BS$_{\text{out}}$ is
present before 80 ns and then removed at 80 ns. In this case, the
$\alpha$ in Eq. (\ref{Psi}) is about $57.3^{\circ}$, which is
determined by the ratio of the coincidence counts for 0-80 ns
($|e\rangle$) and 80-400 ns ($|l\rangle$). The measured
coincidence counts (time-bin 1 ns) at $D_{0}$ is shown in
Fig.2(a), where $\varphi$ are altered from $0^{\circ}$ to
$180^{\circ}$. Before BS$_\text{out}$ is removed (0-80 ns), an
interference fringe determined by $\varphi$ is observed, which
clearly reveals the wave nature of the photon. After
BS$_\text{out}$ is removed (80-400 ns), coincidence counts remain
stable for all $\varphi$ since only one path could be detected by
$D_{0}$, which exhibits the "which-path" information and reveals
the particle nature of the photon. These results are in excellent
agreement with theoretical predictions. More interestingly, the
temporal wave packet of a single photon shown in Fig.2(a)
demonstrates the simultaneous wave and particle behaviors. To our
best knowledge, we directly observe, for the first time, the
simultaneous wave and particle behaviors in one same temporal wave
packet, which will shed light on further understanding of the
wave-particle duality.

The produced wave-particle supposition state can be further
analyzed by choosing different measurement bases in the time bin
degree. As shown in Fig.2(b), in which data are extracted from
Fig. 2(a), if $|e\rangle$ is chosen as the basis, a perfect
interferometer fringe is obtained (red filled circles and line),
and the interference visibility, defined as the ratio of the
oscillation amplitude to the sum of the maximum and minimum
probabilities, is $0.968$. If $|l\rangle$ is chosen as the basis,
a straight line is obtained (blue prisms and line), and the
visibility is 0.043. If $|e\rangle+|l\rangle$ is chosen as the
basis, a wave-particle superposition can be obtained, and the
interference fringe is still observable but with a smaller
visibility of $0.306$ (green triangles and line). Therefore,
Fig.2(b) clearly demonstrates that the visibility in the pure wave
(particle) state is almost one (zero), whereas in the intermediate
cases, the visibility reduces but does not vanish. This result
fits very well with the theoretical prediction and  has been
observed in the quantum delayed-choice experiments\cite{11}.

\begin{figure}
\begin{center}
\includegraphics[width=8cm]{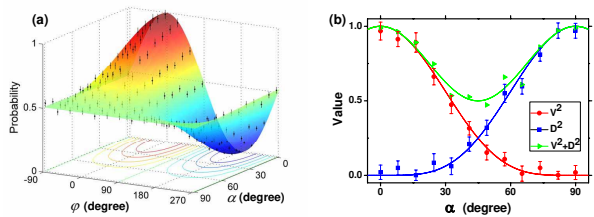}\label{fig:4}
\caption{Characterization of the continuous transition between
wave and particle behaviors. (a) The probability of detecting a
photon at $D_{0}$ as a function of the angles $\alpha$ and
$\varphi$. Dots and error bars represent experimental data points
and their corresponding standard deviations. The experimental data
are fitted by using Eq.(\ref{ID0}) and show excellent agreement
with theoretical predictions. (b) Plots and related fits (solid
lines) of the fringe visibility $V^{2}$ (red filled circles) and
path distinguishability $D^{2}$ (blue squares) as a function of
the angle $\alpha$. For all angles $\alpha$, the inequality
$V^2+D^2 \leq 1$ is verified.}
\end{center}
\end{figure}

In order to further study the features of the above wave-particle
superposition state,  we measure the entire coincidence counts at
$D_{0}$  and then normalized ($I_{D_{0}}$) for $\alpha \in
[0,90^{\circ}]$ and $\varphi \in [-90^{\circ},270^{\circ}]$. As
shown in Fig.3(a), the experimentally measured results are in
excellent agreement with the theoretical predictions of Eq.
(\ref{ID0}). For the angle $\alpha=0^{\circ}$, a perfect
interference fringe with the visibility around one is shown as a
function of $\varphi$, which corresponds to wavelike behavior. For
$\alpha=90^{\circ}$, the measured $I_{D_{0}}=1/2$ and is
independent of $\varphi$, which shows particle-like behavior. For
$0^{\circ}<\alpha<90^{\circ}$, a continuous transition from wave
to particle behavior is observed, which is expressed by the
continually reducing fringe visibility.

As outlined in Ref. \cite{22,23,24,25,26}, a generalized Bohr's
complementarity principle supports the simultaneous observation of
the wave and particle behavior, but the total wavelike and
particle like information (interference fringe visibility $V$ and
path distinguishability $D$) should be limited by the
Englert-Greenberger inequality $V^{2}+D^{2}\leq1$. In our
experiment, $V$ can be obtained from the data shown in Fig.3(a),
whereas $D=\frac{|N_{1}-N_{2}|}{N_{1}+N_{2}}$ should be measured
by blocking one of the beams in the Mach-Zehnder
interferometer\cite{25}. Here $N_{1}$ is the total counts on
$D_{0}(D_{1})$ by blocking one of the beams, and $N_{2}$ is also
the total counts on $D_{0}(D_{1})$ by blocking the other beam. The
results are shown in Fig.3(b), $V^{2}+D^{2}<1$ when
$0^{\circ}<\alpha<90^{\circ}$, $V^{2}+D^{2}=1$ only when
$\alpha=0^{\circ}$ or $\alpha=90^{\circ}$. These results fit very
well with the theoretical predictions and fulfill the
Englert-Greenberger inequality for all angles of $\alpha$.

Although our experimental setup is the same as the classical
delayed-choice experiment in Ref.\cite{9}, the quantum behaviors
as shown in Figs. 2 and 3 are clearly observed in our experiment.
So our experiment suggests that a further generalization of the
light's wave-particle duality is required. Taking a spin half
particle as an example, one can create a superposition state of
spin up and down, and then observe the quantum behaviors of spin
up and down with a classical measurement setup. Similarly, by
employing the language used by Ionicioiu and Terno\cite{10}, one
can write  the wave function of a photon's wave or particle state,
as well as the superposition state of them. The wave-particle
character of the photon  can be considered an additional degree of
freedom, in much the same way as polarization, spin, momentum, and
so on \cite{14}. Therefore, after creating the superposition state
of the wave and particle, it is a predictable result that we can
observe the simultaneous wave and particle behaviors with properly
designed classical measurement setup.  Thus our observation brings
new meaning to the concept of wave-particle duality.

Before ending the manuscript, we briefly address that the approach
developed in our experiment is a convenient way to manipulate the
wave-particle state, which has exciting potential given that the
wave-particle superposition state can be used as a qubit for
quantum information processing\cite{14}. In addition to that, the
parameters $\alpha$ and $\varphi$ in Eq.(\ref{Psi}) are
controllable in our experiment, and the phase $\gamma$ in
Eq.(\ref{Psi}) can be manipulated too. By adding another EOM after
the existing EOM, the phase $\gamma$ between the wave and particle
states can be tunable. In addition, with another unequal arm
Mach-Zehnder interferometer, the time-bin information can be
erased and the phase $\gamma$ can be measured. Furthermore, by
adding a $\lambda/4$ wave plate after the EOM, the wave-particle
supposition can be reversed to
$|\Psi\rangle_{s}(\alpha,\varphi,\gamma)=\cos\alpha|e\rangle|\psi\rangle_\text{p}
+e^{i\gamma}\sin\alpha|l\rangle|\psi\rangle_\text{w}.$

In conclusion, we have  observed the quantum wave-particle
behaviors in a single photon's wave packet with the inserting or
removing of the second BS  through a classical setup. The
utilization of a narrowband single photon enables us to observe
the wave-particle supposition state as  shown in  quantum
delayed-choice experiments but with a classical detecting setup.
The observed results will be useful in further understanding the
light's wave-particle duality. Our experiment can also provide a
feasible new way to create wave-particle superposition state,
which could be useful in quantum information processing.

 This work was supported by the NSF of China (Grants No.
11474107, 11104085, 11125417, and 61378012), the Major Research
Plan of the NSF of China (Grant No. 91121023), the  NFRPC (Grants
No. 2011CB922104 and No. 2013CB921804), the FOYTHEG (Grant No.
Yq2013050), the PRNPGZ (Grant No. 2014010), and the PCSIRT (Grant
No. IRT1243). H. Yan and K. Y. Liao contributed equally to this
work.

\newpage

\begin{widetext}

\begin{center}

\textbf{Supplemental Material for\\
Experimental Observation of Simultaneous Wave and Particle
Behaviors in a Narrowband Single Photon's Wave Packet}

\end{center}

\end{widetext}

\section{Heralded narrowband single photon source.}

Narrowband photon pairs generated through spontaneous four wave
mixing and slow light technique with cold atoms are utilized to
produce the heralded narrowband single photons \cite{15} used in
our experiment. The source is run periodically with a
magneto-optical trap for  trapping time of 4.5 ms and a biphoton
generation time of 0.5 ms. The neutral $^{85}$Rb atoms with an
optical depth of $45$ is trapped in 4.5 ms\cite{19,cpl}. A
four-energy-level double-$\Lambda$ system is chosen for the
spontaneous four wave mixing. In the presence of the pump
($I_{p}\sim 50\mu$W) and coupling lasers ($I_{c}\sim 1.6$ mW), the
counter-propagating Stokes  and anti-Stokes  photons are generated
into opposite directions. The detecting of one Stokes photon,
which also determines the start point of the experiment, heralds
the generation of one narrowband single photon (anti-Stokes
photon). The biphoton generation rate is $47230\ {\text{s}^{ -
1}}$ after taking into account all of the loses. The normalized
cross-correlation function $g^{(2)}_{s,as}(\tau)$ between Stokes
and anti-Stokes photons is around 39, which indicates the
violation of the Cauchy-Schwartz inequality by a factor of 381.
The related conditional second-order correlation of the heralded
single photon $g^{(2)}_{c}=0.23 \pm 0.05$, which indicates a real
single photon.

\section{The controllable polarization beam
splitter (BS$_\text{out}$).}
 In our experimental setup shown in Fig. 1, two polarization beam splitters
(BS$_\text{out}^1$ and BS$_\text{out}^2$) and an electro-optical
modulator (EOM, Newport Model $4102$NF) are combined into a
controllable BS$_\text{out}$ \cite{9}. BS$_\text{out}^1$ combines
the two beams in the Mach-Zehnder interferometer in space, but
they can still be identified by their polarizations. The EOM is
used as a controllable half-wave plate. The axis of the EOM is
aligned $25.5^{\circ}$ to the input polarizations. When $V_{\pi}$
is applied to the EOM, the EOM is equivalent to a half-wave plate
and can rotate the input polarization state (Horizontal and
Vertical) by $45^{\circ}$. In this case, BS$_\text{out}^2$ after
the EOM can mix the two orthogonal polarizations and erase the
path information. This condition occurs when the controllable
BS$_\text{out}$ is present and the interferometer is closed. When
$V=0$ is applied to the EOM, the EOM has no effect on the input
polarization state. In this case, BS$_\text{out}^2$ is simply used
to split the two orthogonal polarizations and the path information
is kept. This condition occurs when the controllable
BS$_\text{out}$ is absent and the interferometer is open. The EOM
voltage between $V_{\pi}=198$ V and $V = 0$ can be switched via a
fast MOSFET (Infineon: BSC16DN25NS3) with a switch off speed
faster than 15 ns. Compared with the total 400 ns temporal length
of the single photon, the maximum influence of the switch off
process to the test probability is less than $0.04$.

\end{document}